\documentclass[12pt]{article}
\usepackage{color}
\usepackage{latexsym}
\usepackage{amsmath}
\usepackage{amsfonts}
\usepackage{amssymb}
\usepackage{indentfirst} 
\numberwithin{equation}{section}
\newcommand{\be}{\begin{equation}}
\newcommand{\ee}{\end{equation}}

\newcommand{\mC }{{\mathbb C}}
\newcommand{\re}{\operatorname{Re}}
\newcommand{\im}{\operatorname{Im}}

\newcommand{\wt}{\omega t}

\title{ Hamiltonian formalisms and symmetries of the Pais-Uhlenbeck oscillator}
\author{Krzysztof  Andrzejewski
\\ \\
\small Department of  Computer Science, \\
\small University of \L\'od\'z,\\
\small Pomorska 149/153, 90-236 {\L}\'od\'z, Poland\\
\small E-mail: k-andrzejewski@uni.lodz.pl
}
\date{}
\begin{document}
\maketitle 
\begin{abstract}
The study of the symmetry of  Pais-Uhlenbeck oscillator initiated in [Nucl. Phys. B 885 (2014) 150] is continued with special emphasis put on the Hamiltonian formalism.  The  symmetry generators  within the original Pais and Uhlenbeck Hamiltonian approach as well as    the canonical transformation    to the  Ostrogradski  Hamiltonian framework are derived.  The resulting algebra of generators appears to be   the central extension of the one obtained on the Lagrangian level; in particular, in the case of odd frequencies one obtains the  centrally extended   $l$-conformal Newton-Hooke algebra. In this important case the canonical transformation to an alternative Hamiltonian formalism (related to the free higher derivatives theory)  is constructed. It is   shown that all generators can be expressed  in terms of the ones for the  free theory   and  the result  agrees with that obtained by  the orbit method. 
\end{abstract}
\section{Introduction}
The theories we are usually dealing with are Newtonian in the sense that the Lagrangian
function depends on the  first time derivatives only. There is, however, an important
exception. It can happen that we are interested only in some selected degrees of
freedom. By eliminating the remaining degrees one obtains what is called an effective
theory. The elimination of a degree of freedom results in increasing the order of dynamical
equations for remaining variables. Therefore,  effective theories are described
by Lagrangians containing higher order  time derivatives  \cite{b1}.  Originally, these theories  were proposed as a
method for dealing with ultraviolet divergences \cite{b2}; this idea appeared to be
quite successful in the case of gravity: the Einstein action supplied by the terms
containing higher powers of curvature leads to a renormalizable theory \cite{b3}. Other examples of higher derivatives theories include
the theory of the radiation reaction \cite{b4,b4a}, the  field theory on noncommutative spacetime
\cite{b5,b5a}, anyons \cite{b6,b6a} or string theories with the  extrinsic curvature \cite{b7}.
\par 
Of course, the appearance of  terms with  higher time derivatives  leads to some problems. One of them is that  the energy  does not need to  be  bounded from below.  
To achieve  a deeper insight into these problems and, possibly, to find a solution it is instructive to consider a  quite  simple, however nontrivial,  higher derivatives theory. For example, it was shown in Ref.  \cite{b8} (see also \cite{b8a})  that  the problem of the energy  can be avoided (on the quantum level) in the case of the celebrated Pais-Uhlenbeck (PU) oscillator \cite{b9}. 
This model has been attracting considerable interest  throughout the years (for  the last few  years, see, e.g., \cite{b8,b8a},\cite{b10a}-\cite{ b11}). 
Recently, it has  been  shown (see,   \cite{b11})  that   the properties of the   PU oscillator, rather surprisingly, for some special values of frequencies change drastically
and  are   related to nonrelativistic conformal symmetries.
Namely,   if  the frequencies of oscillations are {\it odd} multiplicities of a  basic one, i.e., they form  an arithmetic sequences $\omega_k=(2k-1)\omega,\quad  \omega\neq 0$, for $k=1,\ldots,n$, then  the maximal group of Noether symmetries of the PU Lagrangian is the  $l$-conformal Newton-Hooke group  with $l=\frac{2n-1}{2}$ (for more informations about these groups see, e.g., \cite{b12a}-\cite{b12d} and the references therein). Otherwise, the symmetry group is simpler (there are no counterparts of  dilatation and conformal generators (see, the algebra  (\ref{e5})).   
\par
Much attention has been also paid to  Hamiltonian formulations of the PU oscillator. There exists a few approaches to Hamiltonian formalism	 of the PU model: decomposition into the set of the  independent harmonic oscillators proposed by Pais and Uhlenbeck in their original paper \cite{b9}, Ostrogradski approach based on  the Ostrogradski method \cite{b13} of constructing Hamiltonian formalism for theories with higher time  derivatives  and  the last one, applicable in the case of  odd frequencies (mentioned above), which exhibits the   $l$-conformal Newton-Hooke group structure of the model. 
 Consequently, there arises a natural question about the relations between them as well as the realization of the symmetry on the Hamiltonian level? The aim of this work is to give the answer to this question.
 \par 
 The paper is organized as follows.
 After  recalling, in Section 2, some informations concerning symmetry of the PU model on the Lagrangian  level, we start with the  harmonic decoupled approach. We find, on the Hamiltonian level,  the form of generators (for both generic and odd  frequencies)  and we show  that they, indeed, form the algebra which is central extension the one appearing on the Lagrangian level. Section 4 is devoted to the  study of the relation between the above approach and the   Ostrogradski one. Namely, we  construct the canonical transformation which relates the Ostrogradski Hamiltonian to the one describing  the decouple harmonic oscillator.  This transformation enables us to  find the remaining  symmetry generators in terms  of Ostrogradski variables. The next section is devoted to the case of odd frequencies where the additional  natural approach can be constructed. In this framework  the  Hamiltonian  is the sum of the one for  the free higher derivatives theory and  the conformal generator. We derive a  canonical transformation which relates this new Hamiltonian to the one for the  PU oscillator with  odd frequencies. 
Moreover, we apply  the  method (see,  \cite{b14}) of constructing integrals of motion for the systems with symmetry    to  find all symmetry generators. Next,  by direct calculations we show that they are related by the, above  mentioned, canonical transformation to  the ones of the PU model described  in terms decoupled oscillators. We also   express symmetry generators in terms of their counterparts in the free theory.  
\par
 In  concluding Section 6, we summarize our results and  discuss possible further developments.  
 Finally, Appendix constitutes  technical support for the mains results. We derive there some relations and identities which are crucial for our work.   
             
\section{PU oscillator and its symmetry}
Let us consider the three-dimensional   PU oscillator, i.e., the system which is described by the following Lagrangian \cite{b9}
\be
\label{e1}
L=-\frac{1}{2}\vec x \prod_{k=1}^{n}\left(\frac{d^2}{dt^2}+\omega_k^2\right)\vec x, 
\ee
where $0<\omega_1<\omega_2<\ldots<\omega_n$ and $n=1,2,\ldots$.   
Lagrangian  (\ref{e1})  implies the  following equation of motion 
\be
\label{e2}
\prod_{k=1}^n\left(\frac{d^2}{dt^2}+\omega_k^2\right)\vec x=0,
\ee
which  possesses the general solution of the form 
\be
\label{e3}
 \vec x(t)=\sum_{k=1}^{n}(\vec \alpha_k\cos\omega_kt+\vec \beta_k\sin\omega_kt),
\ee
where $\vec \alpha's$ and $\vec \beta's$ are some arbitrary constants.
\par As it has been mentioned  in the Introduction the structure of the maximal symmetry group of Lagrangian (\ref{e1}) depends on the values of $\omega's$.  If  the frequencies of oscillation are { odd}, i.e., they form  an arithmetic sequence $\omega_k=(2k-1)\omega,\quad  \omega\neq 0$,  $k=1,\ldots,n$, then  the maximal group of Noether symmetries of the system  (\ref{e1}) is the $l$-conformal Newton-Hooke group, with $l=\frac{2n-1}{2}$. It is the group which  Lie algebra is spanned by $H,D,K,J^{\alpha\beta}$ and $C_p^\alpha$, $\alpha,\beta=1,2,3$, $p=0,1,\ldots,2n-1$, satisfying the following  commutation rules 
\be
\begin{split}
\label{e4}
[H,D]&=H-2\omega^2 K,\quad  [H,K]=2D,\quad [D,K]=K,\\
[D,\vec C_p]&=(p-\frac{2n-1}{2})\vec C_p,\quad [K,\vec C_p]=(p-2n+1)\vec C_{p+1},\\
[H,\vec C_p]&=p\vec C_{p-1}+(p-2n+1)\omega^2\vec C_{p+1},\\
[J^{\alpha\beta},J^{\gamma\delta}]&=\delta^{\alpha\delta}J^{\gamma\beta}+\delta^{\alpha\gamma}J^{\beta\delta}+\delta^{\beta\gamma}J^{\delta\alpha}+\delta^{\beta\alpha}J^{\alpha\gamma},\\
[J^{\alpha\beta},C_p^\gamma]&=\delta^{\alpha\gamma}C_p^\beta-\delta^{\beta\gamma}C_p^\alpha.
\end{split}
\ee
Although this algebra is isomorphic to the $l$-conformal Galilei one (the latter can be obtained by a linear change of the basis $H\rightarrow H-\omega^2K$, see  \cite{b12a,b12b,b12d} and  \cite{b15a}-\cite{b15f} for more recent developments of this algebra) the use of the basis (\ref{e4}) implies the change of the Hamiltonian which alters the dynamics. 
\par 
In the case of {\it generic} frequencies the maximal symmetry group is simpler. Its Lie algebra consists of $H,J^{\alpha\beta}$ and $\vec C^{\pm}_k$, $k=1,\ldots,n$. The action of $J^{\alpha\beta}$ remains unchanged  and  only commutations  rules between  $H$  and $\vec C's$ must be modified
\be
\label{e5}
\begin{split}
[H,\vec C_k^+]&=-\omega_k\vec C_k^-,\\
[H,\vec C_k^-]&=\omega_k\vec C_k^+.
\end{split}
\ee
\par
 Both symmetry algebras posses central extension:
\be
\label{e6}
[C_p^\alpha,C_q^\beta]=(-1)^pp!q!\delta_{\alpha\beta}\delta_{2n-1,p+q},
\ee
in the  odd case and 
\be
\label{e7}
[C_k^{+\alpha}C_{j}^{-\beta}]=\frac{\omega_k}{\rho_k}\delta_{kj}\delta^{\alpha\beta},
\ee
in the generic case; which will turn out to be necessarily (see the next section) to construct the symmetry algebra on the Hamiltonian level.
\section{Decoupled  oscillators approach}
An approach to the Hamiltonian formalism  of the PU model  was proposed  in Ref.  \cite{b9} where it was demonstrated that the Hamiltonian of the PU oscillator (in dimension one) turns into
 the  sum of the harmonic Hamiltonians with alternating sign. To show this we follow the reasoning of Ref. \cite{b9} and  introduce new variables
\be
\label{e8}
\vec x_k=\Pi_k\vec x, \quad k=1,\ldots,n;                                                 
\ee
where $\Pi_k$ is the projective operator:
\be
\label{e9}
\Pi_k=\sqrt{|\rho_k|}\prod_{\substack{i=1 \\ i\neq k}}^n\left(\frac{d^2}{dt^2}+\omega_i^2\right) ,   
\ee
and
\be 
\label{r0}
\rho_k=\frac{1}{\prod\limits_{\substack{i=1 \\ i\neq k}}^n(\omega_i^2-\omega_k^2)},\quad k=1,2,\ldots,n.
\ee
Note that $\rho_k$ are alternating in sign.
Then one finds
\be
\label{e10}
\vec x=\sum_{k=1}^n(-1)^{k-1}\sqrt{|\rho_k|}\vec x_k,
\ee 
as well as 
\be
\label{e11}
L=-\frac{1}{2}\sum_{k=1}^{n}(-1)^{k-1}\vec x_k\left(\frac{d^2}{dt^2}+\omega_k^2\right)\vec x_k=\frac{1}{2}\sum_{k=1}^{n}(-1)^{k-1}(\dot{\vec{x_k}}-\omega_k^2\vec x_k^2)+t.d.
\ee
The corresponding Hamiltonian reads 
\be
\label{e12}
H=\frac{1}{2}\sum_{k=1}^{n}(-1)^{k-1}(\vec p_k^2+\omega_k^2\vec x_k^2),
\ee
while the canonical  equations of motion are of  the form
\be
\label{e13}
\dot {\vec x}_k=(-1)^{k-1}\vec p_k,\quad \dot{\vec p}_k=(-1)^k\omega_k^2 \vec x_k.
\ee
Taking into account the form of  the general solution (\ref{e3}) we see that the dynamics of the  new canonical variables is given by
\be
\begin{split}
\label{e14}
\vec x_k=\frac{(-1)^{k-1}}{\sqrt{|\rho_k|}}(\vec \alpha_k\cos(\omega_kt)+\vec \beta_k\sin(\omega_kt)),\\
\vec p_k=\frac{\omega_k}{\sqrt{|\rho_k|}}(\vec \beta_k\cos(\omega_kt)-\vec \alpha_k\sin(\omega_kt)).
\end{split}
\ee  
Therefore, we have a correspondence  between the set of solutions of the  Lagrange equation (\ref{e2}) and the set of  solutions of the canonical equations (\ref{e13}). Consequently, we can translate the action of the  group symmetry from the Lagrangian level to the Hamiltonian one and  find all the  symmetry  generators   in terms of oscillator canonical variables. We will show that the generators, obtained in this way,  form the algebra which is the  central extension of the symmetry algebra on the  Lagrangian level.
\par   In the generic case it  is very easy to find  the form of the remaining   (the Hamiltonian is given by (\ref{e12})) symmetry  generators   on the Hamiltonian level. First, let us note that the infinitesimal action of $\vec\mu _k\vec {C}^{+}_k$ and $\vec\nu _k\vec {C}^{-}_k$, $k=1,\ldots,n$, on the Lagrangian level, takes the form
\be
\label{e15}
\begin{split}
\vec x'(t)=\vec x(t)+\sum_{k=1}^n(\vec \mu_k\cos \omega_k t+\vec\nu_k \sin\omega_k t).\
\end{split}
\ee  
Acting with  $\Pi_k$ and applying Eq. (\ref{e13}) we  find   the infinitesimal action of $\vec {C}^{\pm}_k$ on the phase space; by virtue of
 \be
 \label{e16}
 \delta(\cdot)=\epsilon \{\cdot,  \textrm{Generator}\},
 \ee
  we obtain the following generators:
\be
\begin{split}
\label{e17}
\vec C_k^+&=\frac{(-1)^{k-1}}{\sqrt{|\rho_k|}}\cos(\omega_k t)\vec p_k+\frac{\omega_k}{\sqrt{|\rho_k|}}\sin(\omega_k t)\vec x_k,\\
\vec C_k^-&=\frac{(-1)^{k-1}}{\sqrt{|\rho_k|}}\sin(\omega_k t)\vec p_k-\frac{\omega_k}{\sqrt{|\rho_k|}}\cos(\omega_k t)\vec x_k,
\end{split}
\ee
which commute to the central charge -- according to  (\ref{e7}). 
Similarly, the  angular momentum generators read
\be
\label{e18}
J^{\alpha\beta}=\sum_{k=1}^n( x^\alpha_k p_k^\beta-p_k^\alpha x_k^\beta ).
\ee
 Consequently, we obtain  the centrally extended algebra (\ref{e5}).
 \subsection{Odd frequencies} 
In the odd case the symmetry group is reacher  and, therefore, this case is much more interesting. We assume  now that the frequencies form the arithmetic sequence, i.e., $\omega_k=(2k-1)\omega$, $k=1,\ldots,n$. In this case  the main point is that the numbers $\rho_k$ can be explicitly computed; the final result reads 
\be
\label{r1}
\rho_k=\frac{(-1)^{k-1}(2k-1)}{(4\omega^2)^{n-1}(n-k)!(n+k-1)!}, \quad k=1,\ldots,n.
\ee
Consequently, one has  useful  relations
\be
\label{r2}
\frac{|\rho_k|}{|\rho_{k+1}|}=\frac{(2k-1)(n+k)}{(2k+1)(n-k)},\quad k=1,\ldots,n-1.
\ee
Next, let us note  that   the following Fourier expansion holds (see, Appendix)
\be
\label{e20}
\sin^p \wt\cos^{2n-1-p}\wt =
\left\{
\begin{array}{c}
\sum\limits_{k=1}^m\gamma^+_{kp}\cos(2k-1)\wt,  \quad p \textrm{ - even};\\
\sum\limits_{k=1}^m\gamma^-_{kp}\sin(2k-1)\wt,  \quad  p \textrm{ - odd}; 
\end{array}
\right.
\ee
where  $\gamma^\pm_{kp}$ can be expressed in terms of sum of products of binomial coefficients; however,  their  explicit form is not very useful;  for our purposes some properties of $\gamma^\pm_{kp}$ (see,  (\ref{a2})-(\ref{a6}))  will turn out to be  more fruitful.
Now, using Eq. (\ref{e20}) we can rewrite   the infinitesimal action (\ref{e15}), in the case of odd frequencies, in the equivalent form 
\be
\label{e19}
\vec x'(t)=\vec x(t)+\frac{1}{\omega^p}\vec \epsilon_p\sin^p\wt\cos^{2n-1-p}\wt,
\ee  
which gives suitable  family   of the generators $\vec C_p$, $p=0,1,2,\ldots,2n-1$ on the Lagrangian level, i.e.,  satisfying  commutation rules of the $l$-conformal Newton-Hooke algebra (cf.,   \cite{b11})
\par 
In order to find the action of $\vec C_p$ in the Hamiltonian formalism, we use Eqs.   (\ref{e20})  together with  (\ref{e8}) and (\ref{e13}),  which yields
 \be
 \label{e21}
 \vec x'_k=\vec x_k+\frac{(-1)^{k-1}\vec \epsilon_p}{\omega^p\sqrt{|\rho_k|}}\left\{
 \begin{array}{c}
 \gamma^+_{kp}\cos(2k-1)\wt,\quad p \textrm{ - even};\\
 \gamma^-_{kp}\sin(2k-1)\wt,\quad p \textrm{ - odd};
 \end{array}
 \right.
 \ee
\be
\label{e22}
 \vec p'_k=\vec p_k+\frac{(2k-1)\omega\vec \epsilon_p}{\omega^p\sqrt{|\rho_k|}}\left\{
 \begin{array}{c}
 -\gamma^+_{kp}\sin(2k-1)\wt,\quad p \textrm{ - even};\\
 \gamma^-_{kp}\cos(2k-1)\wt,\quad p \textrm{ - odd}.
 \end{array}
 \right. 
 \ee
  Using Eq. (\ref{e16})  we derive the explicit expression for the generators $\vec C_p$ in terms of the canonical variables
  \be
  \label{e23}
   \vec C_p=
  \sum\limits_{k=1}^{n}\frac{\gamma_{kp}^+}{\omega^p\sqrt{|\rho_k|}}\left( (-1)^{k-1}\cos((2k-1)\wt)\vec p_k+(2k-1)\omega\sin((2k-1)\wt)\vec x_k\right),\\
  \ee
  for $ p$  even, and 
  \be
    \label{e23a}
  \vec C_p=
  \sum\limits_{k=1}^{n}\frac{\gamma_{kp}^-}{\omega^p\sqrt{|\rho_k|}}\left( (-1)^{k-1}\sin((2k-1)\wt)\vec p_k-(2k-1)\omega\cos((2k-1)\wt)\vec x_k\right),
\ee
for $p$ odd.
 Eqs. (\ref{e23}) and (\ref{e23a})  can be inverted to yield $\vec x_k$ and $\vec p_k$ in terms of the generators $\vec C_p$
 \be
 \label{e24}
 \begin{split}
  \vec p_k&=(-1)^{k-1}\sqrt{|\rho_k |}\cos((2k-1)\wt)\sideset{}{''}\sum_{p=0}^{2n-1}\beta^+_{pk}\omega^p\vec C_p\\
  &+(-1)^{k-1}\sin((2k-1)\wt)\sideset{}{'}\sum_{p=0}^{2n-1}\beta^-_{pk}\omega^p\vec C_p ,
  \end{split}
 \ee
  \be
  \label{e25}
 \begin{split}
  \vec x_k&=\frac{\sqrt{|\rho_k |}}{(2k-1)\omega}\sin((2k-1)\wt)\sideset{}{''}\sum_{p=0}^{2n-1}\beta^+_{pk}\omega^p\vec C_p\\
 &-\frac{\sqrt{|\rho_k |}}{(2k-1)\omega}\cos((2k-1)\wt)\sideset{}{'}\sum_{p=0}^{2n-1}\beta^-_{pk}\omega^p\vec C_p ,  
  \end{split}
 \ee
 where $\beta^+,\beta^-$  are the inverse matrices to  $\gamma^+,\gamma^-$ while one and two primes $',''$ denote the sum over odd and even indices, respectively\footnote{We will use this convention throughout  the article.}.
 \par
 Next, we find the action of the  dilatation generator. To  this end let us  recall  (cf., \cite{b11}) that  the infinitesimal  action of dilatation on coordinates is   of the form 
 \be
 \label{e26}
\vec x'(t)=\vec x(t)-\frac{\epsilon}{2\omega}\left((2n-1)\omega\cos(2\wt)\vec x(t)-\sin(2\wt)\dot{\vec x}(t)\right).
 \ee
 Substituting (\ref{e10}) and acting with the projectors  $\Pi_k$  we obtain, due to (\ref{e8}) and (\ref{e13}), the  infinitesimal dilatation transformation on the phase space
 \be
 \label{e27}
 \begin{split}
 \vec x_k'&=\vec x_k+\frac{\epsilon}{2\sqrt{|\rho_k|}}\cos(2\wt)\left(\sqrt{|\rho_{k-1}|}(n-k+1)\vec x_{k-1}+\sqrt{|\rho_{k+1}|}(n+k)\vec x_{k+1}\right) \\
 &+\frac{\epsilon(-1)^{k}}{2\omega\sqrt{|\rho_k|}}\sin(2\wt)\left(\frac{\sqrt{|\rho_{k-1}|}}{2k-3}(n-k+1)\vec p_{k-1}- \frac{\sqrt{|\rho_{k+1}|}}{2k+1}(n+k)\right)\vec p_{k+1},\\
\vec x_1'&=\vec x_1-\frac{\epsilon}{2\sqrt{|\rho_1|}}\left(\sqrt{|\rho_{2}|}(n+1)\cos(2\wt)\vec x_{2}+\sin(2\wt)\sqrt{|\rho_{2}|}\frac{(n+1)}{3\omega}\vec p_{2} \right.\\
&\left. -n\cos(2wt)\sqrt{|\rho_1|}\vec x_1+\frac n\omega\sin(2\wt)\sqrt{|\rho_1|} \vec p_1\right),
 \end{split}
\ee
\be
\label{e28}
 \begin{split}
 \vec p_k'&=\vec p_k-\frac{\epsilon(2k-1)}{2\sqrt{|\rho_k|}}\cos(2\wt)\left(\frac{\sqrt{|\rho_{k-1}|}}{2k-3}(n-k+1)\vec p_{k-1}+ \frac{\sqrt{|\rho_{k+1}|}}{2k+1}(n+k)\vec p_{k+1}\right)\\
 &+\frac{\epsilon\omega(-1)^{k}(2k-1)}{2\sqrt{|\rho_k|}}\sin(2\wt)\left(\sqrt{|\rho_{k-1}|}(n-k+1)\vec x_{k-1}-\sqrt{|\rho_{k+1}|}(n+k)\vec x_{k+1}\right) ,\\
 \vec p_1'&=\vec p_1-\frac{\epsilon}{2\sqrt{|\rho_1|}}\cos(2\wt)\left(-n\sqrt{|\rho_{1}|}\vec p_{1}+ \frac{\sqrt{|\rho_{2}|}}{3}(n+1)\vec p_{2}\right)\\
 &+\frac{\epsilon\omega}{2\sqrt{|\rho_1|}}\sin(2\wt)\left(n\sqrt{|\rho_{1}|}\vec x_{1}+\sqrt{|\rho_{2}|}(n+1)\vec x_{2}\right) ,\\
 \end{split}
\ee
where $k>1$ and, by definition,  we put $\vec x_{n+1}=\vec p_{n+1}=0$. One can check, using Eq.  (\ref{r2}), that (\ref{e27}) and (\ref{e28}) define the infinitesimal canonical transformation generated (according to  (\ref{e16})) by  
\be
\label{e29}
D=\frac{-1}{2\omega}\left(\omega A\cos(2\wt)+B\sin(2\wt)\right),
\ee
where
\be
\label{e30}
\begin{split}
A=&-\sum_{k=1}^n\left(\sqrt{\left|\frac{\rho_{k-1}}{\rho_{k}}\right|}(n-k+1)\vec x_{k-1}+\sqrt{\left|\frac{\rho_{k+1}}{\rho_k}\right|}(n+k)\vec x_{k+1}\right)\vec p_k+ n\vec x_1\vec p_1,\\
B=&-\sum_{k=1}^n(-1)^{k}\frac{n-k+1}{2k-3}\sqrt{\left| \frac{\rho_{k-1}}{\rho_{k}}\right| }(\vec p_k\vec p_{k-1}-(2k-1)(2k-3)\omega^2\vec x_k\vec x_{k-1})\\
&+\frac{1}{2}n(\omega^2\vec x_1^2-\vec p_1^2),
\end{split}
\ee
and, by definition, $\vec x_0=\vec p_0=0$. The meaning of the components   $A$ and $B$ will become more clear in Section 5 (see, (\ref{e50})).
\par Similar calculations can be done for the conformal generator $K$. Namely, the infinitesimal conformal transformation, on the Lagrangian level,  reads
\be
\label{e31}
\vec x'(t)=\vec x(t)-\frac{\epsilon}{2\omega^2}\left((2n-1)\omega\sin(2\wt)\vec x(t)+(\cos(2\wt)-1)\dot{\vec x}(t)\right).
 \ee
 \par
 Substituting $\vec x$ and acting with the projector   $\Pi_k$  we obtain the  infinitesimal conformal  transformation on the phase space and consequently (due to (\ref{e16})) the explicit form of  the generator $K$ 
\be
\label{e32}
K=\frac{1}{2\omega^2}\left(B\cos(2\wt)-\omega A\sin(2\wt)+H\right).
\ee 
Finally, the angular momentum takes the same form as in the generic case 
\be
\label{e33}
J^{\alpha\beta}=\sum_{k=1}^n( x^\alpha_k p_k^\beta-p_k^\alpha x_k^\beta ).
\ee 
\par 
It remains to  verify that obtained generators, indeed,  yield integrals of motion and define the centrally extended  $l$-conformal Newton-Hooke algebra. To this end we need a few identities  which are proven in the Appendix.
 First, we compute  the commutators of  $\vec C$'s and check that they   give the proper central extension.
 The only nontrivial case  is   $[C_p^\alpha, C_q^\beta]$ with $p$ even and $q$ odd (or conversely). We have
 \be
 \label{e34}
 \begin{split}
 [ C_p^\alpha, C_q^\beta]&=\frac{\omega(2\omega)^{2(n-1)}\delta^{\alpha\beta}}{\omega^{p+q}} \sum_{k=1}^n(-1)^{k-1}(n-k)!(n+k-1)!\gamma^+_{kp}\gamma^-_{kq}\\
& =\frac{p!(2n-1-p)!\omega^{2n-1}\delta^{\alpha\beta}}{\omega^{p+q}} \sum_{k=1}^n(-1)^{k-1}\beta^+_{pk}\gamma^-_{kq}\\
&=\frac{p!(2n-1-p)!\omega^{2n-1}\delta^{\alpha\beta}}{\omega^{p+q}} \sum_{k=1}^n\beta^-_{2n-1-p,k}\gamma^-_{kq}=\\
&=\frac{p!(2n-1-p)!\omega^{2n-1}\delta^{\alpha\beta}}{\omega^{p+q}} \delta_{2n-1,p+q}=p!q!\delta_{\alpha\beta}\delta_{2n-1,p+q},
 \end{split}
\ee 
where we use consecutively Eqs.  (\ref{e23}),  (\ref{e23a}),   (\ref{r1}), (\ref{a3}) and  (\ref{a2}). For $p$ odd and $q$ even   we obtain the same result except the extra minus sign. Consequently, we obtain the central extension (\ref{e6}). In order to find the remaining commutators let us note that 
\be
\label{e35}
\begin{split}
[A,B]&=-2H,\\
[B,H]&=2\omega^2A,\\
[A,H]&=-2B.
\end{split}
\ee
The proof of the above relations is straightforward  although tedious and involve the use of (\ref{r2}).   Now, by virtue of Eq. (\ref{e35}), it is  easy  to check that the generators  $H,D,K$ satisfy the first line of equations (\ref{e4}).
\par  Now, we find the adjoint action of  $H,D,K,J^{\alpha\beta}$  on $\vec C_p$. Since the calculations are rather wearisome and lengthy we sketch  only the main points.  
To show that   $ [H,\vec C_p] $ gives proper rule we use the identity  (\ref{a4}).  The case $[D,\vec C_p]$ is  more involved; however, using repeatedly Eqs. (\ref{r2})  and (\ref{a5}) we arrive at the desired result.  Similarly to obtain $[H,\vec C_k]$, first, we  use Eq.      (\ref{r2}) and then Eq.  (\ref{a6}).  Finally, it is easy to compute the commutators involving angular momentum. 
\par Having all the commutation  rules and (\ref{a4})  it is not hard to check that the obtained generators are constants  of motion.  This concludes the proof that, on the Hamiltonian level,  they are  symmetry generators  and form the centrally extended   $l$-conformal Newton-Hooke algebra. 
 \section{Ostrogradski approach} 
Since the PU  oscillator is an example of higher derivatives theory,  it is natural to use the Hamiltonian formalism   proposed by Ostrogradski \cite{b13}. To this end let us expand Lagrangian (\ref{e1}) in the sum of higher derivatives terms (here,   $\vec Q=\vec x$)
\be
\label{e36}
L=-\frac{1}{2}\vec Q \prod_{k=1}^{n}\left(\frac{d^2}{dt^2}+\omega_k^2\right)\vec Q=\frac{1}{2}\sum_{k=0}^n(-1)^{k-1}\sigma_k(\vec Q^{(k)})^2 ,
\ee 
where 
\be
\label{e37}
\sigma_k=\sum_{i_1<\ldots<i_{n-k}}\omega_{i_1}^2\cdots \omega_{i_{n-k}}^2,\quad k=0,\ldots,n; \quad \sigma_n=1.
\ee
It can be shown (by standard reasoning)  that  the following identities  hold
\begin{align}
\label{r3}&\sum_{k=1}^n\rho_k\omega_k^{2m}=0, \qquad m=0,\ldots ,n-2, \\
\label{r4}&\sum_{k=1}^n\rho_k\omega_k^{2(n-1)}=(-1)^{n+1},\\
\label{r5}&\sum_{m=0}^n\sigma_m(-1)^m\sum_{k=1}^n\rho_k\omega_k^{2(r-n+m-1)}=0, \quad r\geq n,
\end{align}
where $\rho_k$ is given by Eq. (\ref{r0}). 
Now, we introduce the Ostrogradski variables
\be
\label{e38}
\begin{split}
 \vec Q_k&=\vec Q^{(k-1)}, \\
 \vec P_k&=\sum_{j=0}^{n-k}\left(-\frac{d}{dt}\right)^j\frac{\partial L}{\partial \vec Q^{(k+j)}}=(-1)^{k-1}\sum_{j=k}^n\sigma_j\vec Q^{(2j-k)},
 \end{split}
\ee
for $k=1,\ldots,n$.
Then the Ostrogradski Hamiltonian takes the  form 
\be
\label{e39}
H=\frac{(-1)^{n-1}}{2}\vec P_n^2+\sum_{k=2}^n\vec P_{k-1}\vec Q_k-\frac 1 2 \sum_{k=1}^n(-1)^k\sigma_{k-1}\vec Q_{k}^2.
\ee 
By virtue of Eqs.   (\ref{e13}) and (\ref{e38}),  for $k=1,\ldots,n$,  we find
\be
\begin{split}
\label{e40}
\vec Q_k&=(-1)^{\frac{k-1}{2}}\sum_{j=1}^n\sqrt{|\rho_j|}(-1)^{j-1}\omega_j^{k-1}\vec x_j,\quad k-\textrm {odd};\\
\vec Q_k&=(-1)^{\frac{k}{2}-1}\sum_{j=1}^n\sqrt{|\rho_j|}\omega_j^{k-2}\vec p_j,\quad k- \textrm {even};\\
\end{split}
\ee
and
\be
\label{e41}
\begin{split}
\vec P_k&=(-1)^{\frac k2-1}\sum_{i=1}^n(-1)^{i-1}\sqrt{|\rho_i|}\left(\sum_{j=k}^n\sigma_j(-1)^j\omega_i^{2j-k}\right)\vec x_i, \quad k-\textrm{ even};\\
\vec P_k&=(-1)^{\frac {k-3}{2}}\sum_{i=1}^n\sqrt{|\rho_i|}\left(\sum_{j=k}^n\sigma_j(-1)^j\omega_i^{2j-k-1}\right)\vec p_i, \quad k- \textrm{ odd}.
\end{split}
\ee 
One can show that  Eqs.   (\ref{e40}) and (\ref{e41}) define  a canonical transformation; to compute the Poisson brackets $\{\vec Q_k,\vec Q_j\}$ and   $\{\vec Q_k $, $\vec P_j\}$ we  use (\ref{r3}) and (\ref{r3})-(\ref{r5}), respectively; computing $\{\vec P_k,\vec P_j\}$ is the most complicated one and  involves considering two cases $k-j\lessgtr 1$  as well as  applying Eqs.  (\ref{r3}) and (\ref{r5}).
\par
 Next, let us note that    the inverse transformation is of the form 
\be
\begin{split}
\label{e42}
\vec x_i&=\sideset{}{'}\sum_{k=1}^{n}(-1)^{\frac{k-3}{2}}\sum_{j=k}^n\sigma_j(-1)^j\omega_i^{2j-k-1}\sqrt{|\rho_i|}\vec Q_k+\sideset{}{''}\sum_{k=1}^{n}(-1)^{\frac{k}{2}}\sqrt{|\rho_i|}\omega_i^{k-2}\vec P_k,\\
\vec p_i&=\sideset{}{''}\sum_{k=1}^{n}(-1)^{\frac{k}{2}+i-1}\sum_{j=k}^n\sigma_j(-1)^j\omega_i^{2j-k}\sqrt{|\rho_i|}\vec Q_k+\sideset{}{'}\sum_{k=1}^{n}(-1)^{\frac{k+1}{2}+i}\sqrt{|\rho_i|}\omega_i^{k-1}\vec P_k.
\end{split}
\ee  
\par 
No, we can  try to find the  symmetry generators (both in the odd and  generic  cases) in terms of the Ostrogradski variables. 
Of course,  we  expect that the Hamiltonian (\ref{e12})   should be transformed into the  Ostrogradski one. Indeed, using (\ref{r3})-(\ref{r5}) repeatedly we arrive, after straightforward but rather arduous  computations (considering two cases: $n$ - odd, even),  at the Ostrogradski Hamiltonian (\ref{e39}).
\par Similarly, applying Eqs. (\ref{r3})-(\ref{r5}), we check that the angular momentum (in both cases (\ref{e18}) and (\ref{e33}))  transforms under (\ref{e42})  into Ostrogradski angular momentum
\be
\label{e43}
J^{\alpha\beta}=\sum_{k=1}^n(Q^\alpha_kP^\beta_k-Q^\beta_k P^\alpha_k ).
\ee
 As far as the generators  $\vec C's$ are concerned (again using  (\ref{r3})-(\ref{r5})) we obtain the following expressions:
\be
\label{e44}
\begin{split}
\vec C_k^-=\sum_{k=1}^n(\cos\omega_it)^{(k-1)}\vec P_k-\sum_{k=1}^n\left((-1)^{k-1}\sum_{j=k}^n\sigma_j(\cos\omega_it)^{2j-k}\right)\vec Q_k,\\
\vec C_k^-=\sum_{k=1}^n(\sin\omega_it)^{(k-1)}\vec P_k-\sum_{k=1}^n\left((-1)^{k-1}\sum_{j=k}^n\sigma_j(\sin\omega_it)^{2j-k}\right)\vec Q_k,
\end{split}
\ee
in the case of generic frequencies, and
\be
\label{e45}
\begin{split}
\vec C_p={\frac{1}{\omega^p}}\sum_{k=1}^n\left( \vec P_k(\sin^p\wt\cos^{2n-1-p}\wt)^{(k-1)} \right.\\
\left. +(-1)^k\vec Q_k\sum_{j=k}^n\sigma_j(\sin^p\wt\cos^{2n-1-p}\wt)^{(2j-k)}\right),
\end{split}
\ee
 in the odd case;      
which perfectly agrees with  the definitions of the Ostrogradski canonical variables  (\ref{e38}) and the action of $\vec C's$  on $Q$ (Eqs. (\ref{e15}) and (\ref{e19})).  Similar reasoning can be  done for the remaining two generators $D$ and $K$  in the odd case. Then, they  become  bilinear forms in the Ostrogradski variables; however the explicit form   of coefficients is difficult to simplify and  not  transparent thus we skip it here.  
\section{Algebraic approach to odd case}
Since the    $l$-conformal Newton-Hooke algebra is related to the $l$-conformal Galilei   one by the change of Hamiltonian 
\be 
\label{e46}
H=\tilde H+\omega ^2\tilde  K,
\ee
where tilde refers to generators of  the  free theory (which   possesses the $l$-conformal Galilei symmetry); therefore, it would be   instructive  to construct an alternative Hamiltonian formalism for the PU-model (in the case of {\it odd} frequencies)  with the help of the one for the {\it  free} higher derivatives  theory. 
\par
 Denoting by  $\vec q_m,\vec \pi_m$, $ {m=0},\ldots,{n-1}$ the   phase space coordinates  of the free theory and adapting the results of Ref.  \cite{b16} to our conventions we obtain  the following  form of the  generators of the  free theory (at time $t=0$)
 \be
 \label{e47} 
 \begin{split}
  \tilde H&=\frac{(-1)^{n+1}}{2}\pi_{n-1}^2-\sum_{m=1}^{n-1}\vec q_m\vec \pi_{m-1},\\
 \tilde D&=\sum_{m=0}^{n-1}\left(m-\frac{2n-1}{2}\right)\vec q_m\vec\pi_m,\\
 \tilde K&=(-1)^{n+1}\frac{n^2}{2}\vec q_{n-1}^2+\sum_{m=0}^{n-2}(2n-1-m)(m+1)\vec q_m\vec \pi_{m+1},\\
 \tilde J^{\alpha\beta}&=\sum_{m=0}^{n-1}(q^\alpha_m\pi^\beta_m-q^\beta_m\pi^\alpha_m),\\ 
 \tilde{\vec C}_m&=(-1)^{m+1}m!\vec \pi_m, \quad m=0,\ldots,n-1,\\
 \tilde{\vec C}_{2n-1-m}&=(2n-1-m)!\vec q_m, \quad m=0,\ldots,n-1.
 \end{split}
 \ee
 \par
Of course,  the change of the algebra basis given by   (\ref{e46})  induces the corresponding one for the coordinates  in dual space of the algebra  (denoted  in the same way); consequently  we define the new  Hamiltonian  as follows
\be 
\begin{split}
\label{e48}
H&=\tilde H+\omega^2\tilde K=\frac{(-1)^{n+1}}{2}\pi_{n-1}^2-\sum_{m=1}^{n-1}\vec q_m\vec \pi_{m-1}\\
&+(-1)^{n+1}\frac{n^2\omega^2}{2}\vec q_{n-1}^2+\sum_{m=0}^{n-2}(2n-1-m)(m+1)\omega^2\vec q_m\vec \pi_{m+1}.
\end{split}
\ee 
 We will show that  (\ref{e48})  is indeed  the PU Hamiltonian in $   \vec q_m,\vec \pi_m$ coordinates and we will find the remaining generators in terms of them. To this end  let us define the following transformation 
\be
\label{e49} 
\begin{split}
\vec x_k&=(-1)^k\left(\sideset{}{''}\sum_{m=0}^{n-1}\frac{\omega^{-m}}{m!\sqrt{|\rho_k|}}\gamma^+_{km}\vec q_m+\sideset{}{'}\sum_{m=0}^{n-1}\frac{m!\omega^m\sqrt{|\rho_k|}}{(2k-1)\omega}\beta^+_{2n-1-m, k}\vec \pi_m\right),\\
\vec p_k&=(-1)^{k}\left(\sideset{}{'}\sum_{m=0}^{n-1}\frac{\omega^{-m}(2k-1)\omega}{m!\sqrt{|\rho_k|}}\gamma^+_{k, 2n-1-m}\vec q_m+\sideset{}{''}\sum_{m=0}^{n-1}{m!\omega^m\sqrt{|\rho_k|}}\beta^+_{m k}\vec \pi_m\right),
\end{split}
\ee   
for $k=1,\ldots,n$.
Using (\ref{r2}) and (\ref{a3}) we check that (\ref{e49}) define a canonical transformation. Moreover, by  applying Eqs.  (\ref{r2}) and (\ref{a2})-(\ref{a4})  we check that the PU Hamiltonian (\ref{e12}) (with odd frequencies ) transforms into (\ref{e48}).
The remaining generators can be also  transformed. First,  using (\ref{r2}),  (\ref{a2}), (\ref{a3}), (\ref{a5})  and (\ref{a6}), after  troublesome computations,   we find that  
\be
\label{e50} 
\begin{split}
A&=-2\tilde D,\\
B&=-\tilde H+\omega^2\tilde K,\
\end{split} 
\ee
and, consequently, we obtain a nice interpretation of $A$ and $B$.
Using Eqs. (\ref{e50}), one  checks that $H,D,K$ take the form  
\be
\label{e51} 
\begin{split}
H&=\tilde H+\omega^2 \tilde K,\\
D&=\tilde D \cos 2\wt+\frac{1}{2\omega}(\tilde H-\omega^2\tilde K)\sin 2\wt,\\
K&=\frac 12 (1+\cos 2\wt )\tilde K+ \frac {1}{2\omega^2} (1-\cos 2\wt )\tilde H+\frac {\sin 2\wt }{\omega } \tilde D.\\
\end{split}
\ee
Finally,  the angular momentum reads 
\be
\label{e52} 
J^{\alpha\beta}=\sum_{m=0}^{n-1}(q^\alpha_m\pi^\beta_m-q^\beta_m\pi^\alpha_m) ,
\ee 
i.e., takes the  same form as the one for the  free theory (according to it commutes with $H$). The generators 
 $\vec C_k$ are obtained by plugging  (\ref{e49}) into (\ref{e23}) and (\ref{e23a}),  see also  (\ref{e62}).
\par Summarizing,   we expressed all  PU symmetry  generators  in terms of the ones for free theory (and consequently in terms of $   \vec q_m$ and $\vec \pi_m$) and   
we see that  the  both sets of generators  (except Hamiltonian)  agree at time $t=0$. This result becomes even more evident if we apply  
the algorithm of constructing integrals of motion for Hamiltonian system with symmetry presented in Ref.  \cite{b14}.  Namely, for the Lie algebra spanned by $X_i$, $i=1,\ldots, n$,  $[X_i,X_j]=\sum_{k=1}^nc_{ij}^kX_k$,
with the adjoint action 
\be
\label{e53} 
Ad_g(X_i)=gX_ig^{-1}=\sum_{j=1}^n D^j_i(g)X_j,
\ee
the integrals of motion $X_i(\xi,t)$ corresponding to the generators $X_i$ are of the form
\be
\label{e54} 
X_i(\xi,t)=\sum_{j=1}^nD_i^j(e^{tH})\xi_j,
\ee
where $\xi's$ are the coordinates of the dual space to the Lie algebra (more precisely, their restriction  to the orbits of the coadjoint action  in the dual space).
\par
Let us apply this approach to our case. One can check that    for $H,D,K,J^{\alpha\beta}$  Eq.  (\ref{e54})  gives (\ref{e51}) and  (\ref{e52}).  For $\vec C_p$ we have
\be
\label{e55} 
\vec C_p= e^{tH}\tilde{\vec C}_pe^{-tH}=\sum_{r=0}^{2n-1}a_{pr}(t)\tilde{\vec C}_r, \quad p=0,\ldots,2n-1,
\ee
where the functions $a_{pr}$ satisfy  the set of equations
\be
\label{e56} 
\dot a_{pr}(t)=(r+1)a_{p,r+1}(t)+(r-2n)\omega^2a_{p,r-1}(t),
\ee
with  $a_{k,-1}=a_{k,2n}=0$ and the initial conditions $a_{pr}(0)=\delta_{pr}$ .
Substituting $a_{pr}(t)=\hat a_{pr}(t\omega)\omega^r$ we obtain 
\be
\label{e57}
 \dot {\hat a}_{pr}(t)=(r+1)\hat a_{p,r+1}(t)+(r-2n)\hat a_{p,r-1}(t),
\ee 
with appropriate initial conditions.
It turns out that for fixed $p$  equation (\ref{e57})   is strongly related to the  evolution of ${\vec q}$'s and $\vec \pi$'s  in  the PU model with odd frequencies. More precisely, the canonical equations of motion for the Hamiltonian (\ref{e48})  are equivalent to  Eq. (\ref{e57}) for  fixed $p$ (cf.,  \cite{b17}).   Consequently,  the solution can be written in terms of combinations of harmonics with odd frequencies: 
\be
\label{e58}
\hat a_{pr}(t)=\sideset{}{'}\sum_{a=-(2n-1)}^{2n-1}i^r\beta_{ra}e^{iat}s_{a}^p,
\ee 
where $\beta_{pa}$ is given by (\ref{a23}) and $s^p_a$ are some constants (see, \cite{b17})). Taking into account the initial conditions,  we obtain 
\be
\label{e59}
a_{pr}(t)=\omega^{r-p}\sideset{}{'}\sum_{a=-(2n-1)}^{2n-1}i^r\beta_{ra}\gamma_{ap}e^{iat\omega}.
\ee 
By virtue of Eqs.  (\ref{a24}) and (\ref{a28}), we have
\be
\label{e60}
\begin{split}
a_{pr}(t)=\omega^{r-p}\sum_{k=1}^{n}\beta_{rk}^\pm\gamma_{kp}^\pm \cos {(2k-1)\wt},
\end{split}
\ee
where upper (lower) sign corresponds to $p,r$ even (odd);  and  
\be
\label{e61}
\begin{split}
a_{pr}(t)=\mp\omega^{r-p}\sum_{k=1}^{n}\beta_{rk}^\mp\gamma_{kp}^\pm \sin {(2k-1)\wt},
\end{split}
\ee 
where upper (lower) sign corresponds to $p$ even and  $r$ odd ($p$ odd and  $r$ even). 
Having the explicit  form of  $a_{pr}(t)$, and using  Eqs. (\ref{e47}) and (\ref{e55})   we  obtain $\vec C's$ in terms of $\vec q$ 's and  $\vec \pi$'s:
\be
\label{e62}
\vec C_p=\sum_{r=0}^{n-1}\left((-1)^{r-1}r!a_{pr}(t)\vec \pi_r+(2n-1-r)!a_{p,2n-1-r}(t)\vec q_r\right).
\ee 
As we have mentioned  above   (\ref{e62}) is related by canonical transformation (\ref{e49}) to (\ref{e23}) and (\ref{e23a}).
\section{Discussion}
Let us summarize. In the present paper we focused on the Hamiltonian approaches to the PU model and its symmetries.  First, we derived the form of  the symmetry generators,  in the original Pais and Uhlenbeck approach (for both generic and odd frequencies).  We have shown that the resulting algebra is  the central extension of  the one obtained on the Lagrangian level, i.e., the  centrally extended   $l$-conformal Newton-Hooke  algebra in the case of odd frequencies and the algebra defined by Eqs.  (\ref{e5}) and  (\ref{e7}),  in the generic case.      
Next, we considered the  Ostrogradski  method of constructing Hamiltonian formalism for theories with higher derivatives. We  derived the canonical transformation (Eqs.  (\ref{e40})-(\ref{e41}))  leading the Ostrogradski Hamiltonian  to the one in   decoupled  oscillators approach.
\par
Let us note that the  both approaches, mentioned above,  do not distinguish  the odd   frequencies and in that case do not uncover the  richer symmetry. A deeper insight  is attained by nothing that for odd frequencies an alternative Hamiltonian formalism  can be constructed. It is based on the Hamiltonian formalism   for the free higher derivatives theory exhibiting the $l$-conformal Galilei  symmetry.
More precisely, we add  to the Hamiltonian of the  free theory the conformal generator. As a result, we obtain the  new Hamiltonian, which turns out to be  related, by canonical transformation (\ref{e49}), to the PU one.  This construction  can be better understood from the orbit method point of  view, where the construction of  dynamical realizations of a given symmetry algebra is related to a choice  of one element of the dual space of the algebra as the  Hamiltonian (see, \cite{b14} and the references therein). In our case, both  algebras ($l$-Galilei and $l$-Newton-Hooke) are isomorphic to each other; only  the  one generator,  corresponding to the Hamiltonian, differ by adding the  conformal generator of the  free theory. This  gives the suitable change in the dual space and consequently the definition (\ref{e48}). 
\par
The change of  the Hamiltonian alters the dynamics, which  implies different  time dependence of the symmetry generators   (which do not commute with $H$); however,  all PU generators should  be expressed  in terms of the generators of the  free theory (for $t=0$).  This fact was confirmed by applying   the  method presented in Ref. \cite{b14}  as well as,  directly,  by the canonical transformation (\ref{e49})  to the decoupled  oscillators approach for the PU model.
\par
Turning to possible further developments, let us recall  that  in the classical case ($l=\frac 12$) the dynamics of harmonic oscillator (on the half-period) is related  to the dynamics of free particle by well known   Niederer's transformation \cite{b18} (this fact  has also counterpart on  the quantum level). It turns out that this relation can be generalized to an arbitrary half-integer $l$ \cite{b11} on the Lagrangian level; on the Hamiltonian one, we encounter some difficulties since there is no straightforward 
 transition to the  Hamiltonian formalism for a theory with higher derivatives.  
 However, in the recent paper \cite{b19} the canonical transformation    which relates  the Hamiltonian  (\ref{e48}) to the one for free theory (the first line of (\ref{e47})) has been  constructed; it provides  a counterpart of classical  Niederer's transformation for the  Hamiltonian formalism developed  in Section 5.   Using  our results  one can obtain similar transformation for both  remaining Hamiltonian approaches. We also believe that  the results presented   here  can help in constructing quantum counterpart of the Niederer's transformation for higher $l$ as well as to study of  the symmetry  of the  quantum version of  PU oscillator. 
\vspace{0.5cm}
\par
{\bf Acknowledgments.}
The author is grateful to Joanna  Gonera, Piotr Kosi\'nski and Pawe\l\  Ma\'slanka for useful comments  discussions.  The e-mail discussion with Professors   Anton  Galajinsky and  Ivan Masterov is highly acknowledged.
The work is supported by the grant of National Research Center number 
DEC-2013/09/B/ST2/02205.
 \appendix
\section{Appendix}In this Appendix we prove the following Fourier expansion
\be
\label{a1}
\sin^p t\cos^{2n-1-p}t=
\left\{
\begin{array}{c}
\sum_{k=1}^n\gamma^+_{kp}\cos(2k-1)t, \qquad p \textrm{ - even};\\
\sum_{k=1}^n\gamma^-_{kp}\sin(2k-1)t, \qquad p \textrm{ - odd}; 
\end{array}
\right.
\ee
and derive some, crucial for the main part of the paper,    properties of the  expansion  coefficients; namely
\be
\label{a2}
 \gamma^+_{kp}=(-1)^{k-1}\gamma^-_{k,2n-1-p}, \qquad \beta^+_{pk}=(-1)^{k-1}\beta^-_{2n-1-p,k},
\ee
\\
\be
\label{a3}
2p!(2n-1-p)!\beta^\pm_{pk}=2^{2n-1}(n-k)!(n+k-1)!\gamma^\pm_{kp},
\ee
\\
\be
\label{a4}
(2k-1)\gamma^\pm_{kp}=\mp p\gamma^\mp_{k,p-1}\pm(2n-1-p)\gamma^\mp_{k,p+1},
\ee
\\
\be
\label{a5}
(n+k)\gamma^\pm_{k+1,p}+(n-k+1)\gamma^\pm_{k-1,p}\pm n\gamma^\pm_{kp}\delta_{k1}=(2n-1-2p)\gamma^\pm_{kp},
\ee
\\
\be
\label{a6}
(n+k)\gamma^\pm_{k+1,p}-(n-k+1)\gamma^\pm_{k-1,p}\mp n\gamma^\pm_{kp}\delta_{k1}=\mp p\gamma^\mp_{k,p-1}\mp(2n-1-p)\gamma^\mp_{k,p+1},
\ee
where $\beta^\pm$ is the inverse  matrix of $\gamma^{\pm}$ and by definition  $\gamma^\pm_{kp}=0 $ whenever $p<0, p>2n-1, k<1,k>n$.   
Let us stress that $\beta^+_{pk},\gamma^+_{kp}$ $ (\beta^-_{pk},\gamma^-_{kp})$ are defined only for $p$ even (odd). 
\par   Let us consider,  for fixed $n$, $n=1,2,\ldots$
 the set of functions 
\be
\label{a7}
\begin{split}
P^+_k(\tau)&=\frac{\sqrt 2 \cos(2k-1)t}{\cos^{2n-1}t}\Big|_{t=\arctan\tau}, \\
P^-_k(\tau)&=\frac{\sqrt 2 \sin(2k-1)t}{\cos^{2n-1}t}\Big|_{t=\arctan\tau}, 
\end{split}
\ee
where $k$ is, a priori, an integer.
 One can check that   functions (\ref{a7})  satisfy the orthonormality relations   
\be
\begin{split}
\label{a8}
\int_{-\infty}^{\infty}\frac{P_k^+(\tau)P_j^+(\tau)}{\pi (1+\tau^2)^{2n}}d\tau&=
\int_{-\infty}^{\infty}\frac{P_k^-(\tau)P_j^-(\tau)}{\pi (1+\tau^2)^{2n}}d\tau=\delta_{kj},\\
\int_{-\infty}^{\infty}\frac{P_k^\pm(\tau)P_j^\mp(\tau)}{\pi (1+\tau^2)^{2n}}d\tau&=0 ,
\end{split}
\ee
and    the following identities
\begin{align}
\label{a9}P_0^\pm&=\pm P_1^\pm,\\
\label{a10}(1+\tau^2)(P_k^\pm)'&=\mp(2k-1)P_k^\mp+(2n-1)\tau P^\pm_k,\\
\nonumber\\
\label{a11}(1+\tau ^2)P_{k+1}^\pm&=P_k^\pm(1-\tau^2)\mp 2\tau P_k^\mp,\\
\label{a12}(1+\tau ^2)P_{k-1}^\pm&=P_k^\pm(1-\tau^2)\pm 2\tau P_k^\mp,\\
\nonumber\\
\label{a13}(n-k)P_{k+1}^\pm+(n+k-1)P_{k-1}^\pm&=(2n-1)P_k^\pm-2\tau(P_k^\pm)' ,\\
\label{a14}(n-k)P^\pm_{k+1}-(n+k-1)P^\pm_{k-1}&=(2k-1)P_k^\pm \mp2(P_k^\mp)'.
\end{align}   
Let $X$ denote the operator
\be
\label{a15}
X=(1+\tau^2)\frac{d}{d\tau}-(2n-1)\tau.
\ee
Then 
\be
\label{a16}
XP_k^\pm =\mp(2k-1)P_k^\mp;
\ee 
consequently the action of the  operator $Y=X^2$ is as follows 
\be
\label{a17}
YP_k^{\pm}=-(2k-1)^2P_k^\pm,
\ee
i.e., $P's$ are eigenvectors of the operator $Y$.
\par
 Now, the point is that  for $k=1,\ldots, n$ the functions $P_k^{\pm}$  are  {\it polynomials} of degree less than or equal to $2n-1$ (this can be seen by expanding $\sin(2k-1)t$ and $\cos(2k-1)t$) . Due to (\ref{e8})  they form the orthonormal basis in the space $W^{2n-1}(\tau)$ of all polynomials degree less than $2n$  with the scalar product  
\be
(f,g)=\int_{-\infty}^{\infty}\frac{f(\tau)g(\tau)}{\pi (1+\tau^2)^{2n}}d\tau.
\ee
Since $P_k^+,(P_k^-)$ are  even (odd) functions   the expansion with respect to the standard basis $\{\tau^p\}_{p=0}^{2n-1}$  of $W^{2n-1}(\tau)$  is of the form
\be
\begin{split}
\label{a18}
P_k^+(\tau)=\sqrt{2}\sideset{}{''}\sum_{p=0}^{2n-1}\beta^+_{pk}\tau^p, \quad k=1\ldots n;\\ 
P_k^-(\tau)=\sqrt{2}\sideset{}{'}\sum_{p=0}^{2n-1}\beta^-_{pk}\tau^p,  \quad k=1\ldots n .
\end{split}
\ee
 Moreover, since $P_0^+=P_1^+$ and $P_0^-=-P_1^-$  we have $\beta_{p0}^+=\beta_{p1}^+$ and $\beta_{p0}^-=-\beta_{p1}^-$ Denoting by $\gamma^{\pm}$ the inverse matrix of $\beta^{\pm}$  we get the following  relations
\be
\begin{split}
\label{a19}
\tau ^p&=\frac{1}{\sqrt 2}\sum_{k=1}^n{\gamma^{+}_{kp}}P^+_k(\tau), \quad p - \textrm{even};\\
\tau ^p&=\frac{1}{\sqrt 2}\sum_{k=1}^n{\gamma^{-}_{kp}}P^-_k(\tau), \quad p - \textrm{odd}.\\
\end{split}
\ee
Substituting $\tau=\tan t$ in Eqs.  (\ref{a19}) we obtain  the expansions
\be
\begin{split}
\label{a20}
\tan^p t&=\sum_{k=1}^n{\gamma^{+}_{kp}}\frac{\cos(2k-1)t}{\cos^{2n-1}t}, \quad p - \textrm{even};\\
\tan^p t &=\sum_{k=1}^n{\gamma^{-}_{kp}}\frac{\sin(2k-1)t}{\cos^{2n-1}t}, \quad p - \textrm{odd};\\
\end{split}
\ee
which are equivalent to the Fourier expansion  (\ref{a1}). 
\par Now, we prove the identities (\ref{a2}) -(\ref{a6}). First, let us note that the operator $X$ was considered in Ref.  \cite{b17}\footnote{For our convention   $n$ must be replaced  there with  $n-1 $}  as acting on the space  $W^{2n-1}_\mC$ (the space of complex values polynomials of degree less than $2n$).  It was shown there that 
the  polynomials
\be
\label{a21}
P_a(\tau)=(1+i\tau)^\frac{2n-1+a}{2}(1-i\tau)^\frac{2n-1-a}{2},
\ee 
where the index $a$ is an odd integer  belonging to the set  $\{-(2n-1),\ldots,(2n-1) \}$,
 form an  orthonormal basis of   $W^{2n-1}_\mC$  and  are the eigenvectors   of $X$, i.e., 
\be
\label{a22}
XP_a=iaP_a.
\ee
Moreover, it was proved that the coefficients of the expansion 
\be
\label{a23}
P_a(\tau)=\sum_{p=0}^{2n-1}i^p\beta_{pa}\tau^p,
\ee
satisfy the relations
\be
\label{a24}
\beta_{p,-a}=(-1)^p\beta_{pa}, \quad \beta_{2n-1-p,a}=(-1)^{\frac{2n-1-a}{2}}\beta_{pa}.
\ee
Furthermore, with   $(\gamma_{ap})$ being  the inverse matrix to $(\beta_{pa})$ the following important relation holds    
\be
\label{a25}
p!(2n-1-p)!\beta_{pa}=G(n,a)\gamma_{ap}i^p,
\ee
where 
\be
\label{a26}
G(n,a)=2^{2n-1}\left(\frac{2n-1+a}{2}\right)!\left(\frac{2n-1-a}{2}\right)!.
\ee
\par
We can use this information to obtain some relations for $\beta^\pm$ and $\gamma^\pm$. To this end let us note that we have 
\be
\begin{split}
\label{a27}
P_k^+&=\sqrt2 \re(P_{2k-1}) =\sqrt 2 \re(P_{-(2k-1)}),\\
P_k^-&=\sqrt 2 \im(P_{2k-1}) =-\sqrt 2\im (P_{-(2k-1)}), \\
\end{split}
\ee
which implies
\be
\label{a28}
\begin{split}
&\beta_{pk}^+=(-1)^{\frac{p}{2}}\beta_{p,2k-1}, \quad
\beta_{pk}^-=(-1)^{\frac{p-1}{2}}\beta_{p,2k-1},\\
&\gamma^+_{kp}=2\gamma_{2k-1,p}, \quad
\gamma^-_{kp}=2i\gamma_{2k-1,p} ,
\end{split}
\ee
where $p$ is even (odd) for the  $+(-)$ case, respectively.
\par  Now,  we are ready to prove the  relations (\ref{a2})--(\ref{a6}). First, using (\ref{a24}),  (\ref{a25})  and (\ref{a28}) we get  (\ref{a2}) and (\ref{a3}).
Recursion relation  (\ref{a4}) is  obtained by differentiating (\ref{a1}).  Substituting (\ref{a18}) to (\ref{a13})   and using (\ref{a3}), (\ref{a9}) we arrive at (\ref{a5}). Similarly, inserting    (\ref{a18}) into (\ref{a14})  and applying  (\ref{a2}),  (\ref{a3}), (\ref{a9}) we get (\ref{a6}).

\end{document}